\documentclass[10pt,letterpaper]{article}

\usepackage[utf8]{inputenc}
\usepackage{subfig,eurosym}
\usepackage{physics}
\usepackage{comment}
\usepackage[svgnames]{xcolor}
\usepackage{amsmath,amssymb,amsfonts,amsthm}
\usepackage{mathtools,old-arrows,verbatim,mathrsfs}
\usepackage{graphicx}
\usepackage{epstopdf,epsfig,changepage}

\usepackage{bm}
\usepackage[top=2.5cm, bottom=3cm, left=3.5cm, right=3.5cm,
           heightrounded,marginparwidth=1.5cm, marginparsep=1cm]{geometry}
\usepackage[shortlabels]{enumitem}
\usepackage[square,numbers,merge,comma,sort&compress]{natbib}
\makeatletter
\def\NAT@spacechar{\,}
\makeatother
\usepackage{relsize,setspace,moresize}
\usepackage{latexsym,mathrsfs,calligra,aurical}
\usepackage{float,appendix,xargs,extarrows,empheq,url}
\usepackage{cancel}
\usepackage{jheppub}
\hypersetup{
    pdfmenubar=true,pdffitwindow=false,pdfpagemode={UseNone},
    pdfstartview={FitH},colorlinks=true,bookmarksnumbered=true,
    plainpages,linktoc=page,citecolor=blue,filecolor=black,
    linkcolor=Crimson,urlcolor=Green}
\usepackage{footnotebackref}
\usepackage{hypernat}

\DeclareFixedFont\trfont{OT1}{phv}{b}{sc}{11}
\def\={\:=\:}

\newcommandx{\ETh}[2][1=M,2=\alpha,usedefault]{\Theta_{#1}{}^{#2}}
\newcommandx{\overbar}[1]{\mkern1.5mu\overline{\mkern-2.0mu#1\mkern-2.0mu}\mkern1.5mu}
\newcommandx{\overbarM}[1]{\mkern6.0mu\overline{\mkern-5.5mu#1\mkern-3.5mu}\mkern1.5mu}
\newcommandx{\overbarcal}[1]{\mkern6.0mu\overline{\mkern-5.5mu#1\mkern-1.0mu}\mkern1.5mu}
\DeclareFixedFont\trfont{OT1}{phv}{b}{sc}{11}
\DeclareMathAlphabet{\mathpzc}{OT1}{pzc}{m}{it}
\DeclareMathAlphabet{\mathcal}{OMS}{cmsy}{m}{n}
\DeclareSymbolFontAlphabet{\Scr}{rsfs}
\DeclareMathAlphabet{\mathbold}{U}{BOONDOX-ds}{m}{n}
\SetMathAlphabet{\mathbold}{bold}{U}{BOONDOX-ds}{b}{n}
\DeclareMathAlphabet{\mathcalboondox}{U}{BOONDOX-calo}{m}{n}
\SetMathAlphabet{\mathcalboondox}{bold}{U}{BOONDOX-calo}{b}{n}
\DeclareMathAlphabet{\mathbcalboondox}{U}{BOONDOX-calo}{b}{n}

\title{\centering\boldmath\LARGE\bfseries{%
Holographic renormalization and the variational problem for mixed boundary conditions via a solution-dependent superpotential-like function
}\vspace{1.25em}}

\author[a]{David Choque}
\emailAdd{dchoque@pucp.edu.pe}

\author[b]{Ra\'ul Rojas}
\emailAdd{raul.rojas@uda.cl}

\affiliation[a]{Pontificia Universidad Cat\'olica del Per\'u, Av. Universitaria 1801, San Miguel, Per\'u.}
\affiliation[b]{Departamento de F\'isica, Universidad de Atacama, Avenida Copayapu 485, Copiap\'o, Chile.}

\abstract{
We study holographic renormalization and the variational problem in four-dimensional Einstein gravity coupled to a self-interacting scalar field in asymptotically AdS spacetimes with mixed, designer-gravity boundary conditions. For static black-hole solutions, we introduce a solution-dependent superpotential-like function $W(\phi)$, motivated by the Hamilton--Jacobi formulation but defined directly from the equations of motion. Focusing on the case $m^{2}L^{2}=-2$, we show that the near-boundary expansion
$
W(\phi)=-\frac{4}{L}-\frac{\phi^{2}}{2L}+a\phi^{3}+\mathcal{O}(\phi^{4})
$
is not fully determined by the bulk equations. Instead, once integrable mixed boundary conditions $B=B(A)$ are imposed and the variational principle is required to be well posed, the cubic coefficient is fixed in terms of the boundary deformation. In this way, the mixed boundary condition is encoded directly in the scalar counterterm, rendering the Euclidean on-shell action finite without the need for additional scalar boundary terms. We then derive the renormalized Euclidean action and holographic stress tensor, verify the quantum-statistical relation under mixed boundary conditions, and show that $W(\phi)$ provides a natural characterization of holographic renormalization-group data in non-extremal backgrounds. Finally, we illustrate the formalism in exact asymptotically AdS black-hole solutions arising in consistent truncations, including a case where comparison with a supergravity superpotential clarifies why the RG observables are controlled by the solution-dependent function $W(\phi)$ rather than by $W_{\text{SUGRA}}$.
}

\date{}

\begin{document}

\maketitle

\section{Introduction}

Holographic renormalization provides a systematic framework for rendering the on-shell gravitational action finite and formulating a well-posed variational principle in asymptotically anti--de Sitter (AdS) spacetimes \cite{deBoer:1999tgo,Bianchi:2001kw,Skenderis:2002wp,deBoer:2000cz,Papadimitriou:2004ap}. The divergences arising from the asymptotic region near the AdS boundary are canceled by supplementing the bulk action with appropriate local boundary terms. Within the AdS/CFT correspondence \cite{Maldacena:1997re}, the renormalized on-shell gravitational action acquires a direct interpretation as the generating functional of connected correlators in the dual CFT \cite{Witten:1998qj,Gubser:1998bc,Klebanov:1999tb}. Crucially, this identification depends on the choice of boundary data and on the boundary terms added to the gravitational action. Different boundary conditions therefore correspond to different deformations of the dual field theory, and the renormalized action encodes not only the bulk dynamics but also the precise form of these deformations.

When the admissible boundary data are not uniquely fixed, the construction of boundary terms, and hence the implementation of a consistent variational principle, requires additional care. This situation is especially subtle in Einstein--scalar theories whose scalar mass lies in the Breitenlohner--Freedman window. In that regime, both independent near-boundary fall-off modes are normalizable. In this case, bulk dynamics alone does not select a unique boundary condition, and one may consistently impose mixed boundary conditions relating the leading and subleading modes. These define the class of models known as designer gravity \cite{Papadimitriou:2004ap,Hertog:2004dr}. From the holographic viewpoint, such boundary conditions correspond to multi-trace deformations of the dual CFT \cite{Witten:2001ua,Papadimitriou:2004ap,Hertog:2004dr,Compere:2008us}. Implementing these deformations within holographic renormalization requires a careful treatment of the boundary terms, as the choice of boundary condition determines not only the dual theory but also the precise form of the variational problem and its finite contributions. This choice has important consequences for the spectrum, stability, and thermodynamic properties of the corresponding AdS solutions \cite{Hertog:2005hu}.

Such scalar fields arise naturally in lower-dimensional effective theories obtained from Kaluza--Klein compactifications and from consistent truncations of higher-dimensional supergravity \cite{Duff:1986hr,Green:1987sp,Candelas:1985en,Lu:1998xt}. In asymptotically AdS spacetimes their near-boundary behavior involves two independent normalizable modes, whose relation is not fixed by bulk dynamics. From the holographic perspective, these modes are mapped to the source and expectation value of a scalar operator in the dual CFT, and mixed boundary conditions correspond to specifying a functional relation between them. While this correspondence is well understood at the level of the AdS/CFT dictionary, its implementation within holographic renormalization is subtler. In particular, imposing a general relation between the asymptotic modes requires a careful treatment of finite boundary terms in order to ensure a well-posed variational problem and a consistent definition of the renormalized action.

In supersymmetric theories, such as gauged supergravity, the scalar potential $V(\phi)$ can often be expressed in terms of a superpotential $W_{\text{SUGRA}}$ as \cite{Freedman:1999gp,DeWolfe:1999cp}
\begin{equation} 
V(\phi) = \frac{1}{2} G^{ij}(\phi)\,\partial_i W_{\text{SUGRA}}(\phi)\,\partial_j W_{\text{SUGRA}}(\phi) - \frac{d}{4(d-1)} W_{\text{SUGRA}}^2(\phi), 
\label{eq:intro_sugra_potential}
\end{equation}
where $G_{ij}$ is the metric on the scalar target space and $d$ is the dimension of the boundary theory. Whenever such a relation holds, the second-order equations of motion can be reduced to first-order flow equations, which have been widely used in the construction of domain-wall solutions dual to renormalization-group flows, as well as extremal black holes \cite{Ceresole:2007wx,Skenderis:2006jq}. Even in the absence of supersymmetry, one may sometimes impose an analogous algebraic relation involving a fake superpotential, leading to the framework of fake supergravity \cite{Townsend:1984iu,Skenderis:1999mm}.

However, for generic finite-temperature black-hole solutions, the existence of a globally defined superpotential $W_{\text{SUGRA}}(\phi)$ governing first-order flow equations is not guaranteed. Nevertheless, within the Hamilton--Jacobi formulation of gravitational dynamics, it is well known that along a given solution one may introduce a superpotential-like function, denoted here by $W(\phi)$, which captures the divergent structure of Hamilton's principal function and organizes the radial dependence of the fields.
Unlike $W_{\text{SUGRA}}(\phi)$, $W(\phi)$ is a solution-dependent quantity, defined only along a given background, and is not determined solely by the scalar potential.

Near the AdS boundary, where thermal effects are subleading, the asymptotic expansion for $W(\phi)$ coincides with that of a fake superpotential associated with the AdS critical point. It is precisely this near-boundary expansion that is relevant for the construction of boundary counterterms, the renormalization of the on-shell action, and the formulation of a consistent variational principle, even in non-supersymmetric and finite-temperature settings.

A second motivation for introducing $W(\phi)$ arises from holographic renormalization-group physics. In finite-temperature backgrounds with horizons, the radial evolution is naturally expressed in terms of quantities defined along a given solution, and the superpotential-like function $W(\phi)$ provides the appropriate geometric object for this purpose. In particular, for a running coupling $\lambda\equiv e^{\phi}$, the holographic $\beta$-function can be written as
\begin{equation}
\label{eq:introbeta}
\beta_{\lambda}(\lambda)=-\frac{4}{W(\phi)}\frac{dW(\phi)}{d\phi}\lambda,
\end{equation}
while the holographic $c$-function can likewise be expressed in terms of $W(\phi)$.

For a given black-hole branch, the RG flow is therefore controlled by the solution-dependent function $W(\phi)$, rather than by a superpotential defined solely at the level of the scalar potential. More precisely, one finds that, in general,
\begin{equation}
\label{eq:introineq}
-\frac{4}{W(\phi)}\frac{dW(\phi)}{d\phi}\lambda\neq -\frac{4}{W_{\text{SUGRA}}(\phi)}\frac{dW_{\text{SUGRA}}(\phi)}{d\phi}\lambda,
\end{equation}
reflecting the fact that the two constructions are generically inequivalent. This distinction is especially important at finite temperature, where the RG observables relevant for a given background are determined by the function reconstructed from the geometry itself. When expressed in terms of $W(\phi)$, the holographic $c$-function approaches a constant at the AdS boundary, as expected at a conformal fixed point.

In this work we revisit holographic renormalization in designer gravity, taking the variational principle as the organizing principle throughout. While the Hamilton--Jacobi formulation provides the natural motivation for introducing a superpotential-like function, our construction of $W(\phi)$ is carried out directly at the level of the equations of motion and is tailored to the class of static solutions considered here. We focus on a scalar with $m^{2}L^{2}=-2$, for which
\begin{equation}
\label{eq:introfalloff}
\phi(r)=\frac{A}{r}+\frac{B}{r^{2}}+\mathcal{O}(r^{-3}),
\end{equation}
and show that the near-boundary expansion of the superpotential-like function takes the form
\begin{equation}
\label{eq:introW}
W(\phi)=-\frac{4}{L}-\frac{\phi^{2}}{2L}+a\phi^{3}+\mathcal{O}(\phi^{4}),
\end{equation}
with the cubic coefficient uniquely fixed once integrable mixed boundary conditions are imposed and the variational problem is required to be well posed. Equivalently, writing the boundary condition as $B=B(A)$ with $B(A)=dw(A)/dA$, this requirement determines the asymptotic counterterm constructed from $W(\phi)$, thereby incorporating the boundary deformation directly into the renormalized action. As a result, the scalar counterterm is completely fixed without the need for additional scalar boundary terms beyond the standard geometric contributions.

We then use $W(\phi)$ to derive the renormalized Euclidean action, the quasilocal energy, and the holographic stress tensor under mixed boundary conditions. Moreover, $W(\phi)$ provides a unified description of holographic renormalization-group data in non-extremal backgrounds: it yields a natural holographic $\beta$-function and allows the holographic $c$-function to be expressed in a form that approaches the expected constant value at the AdS boundary. We illustrate the formalism in exact asymptotically AdS black-hole solutions arising from consistent truncations. In particular, we analyze one example directly in terms of the superpotential-like function reconstructed from the geometry, and another in which a supergravity superpotential exists, allowing a direct comparison that clarifies the role of the solution-dependent function $W(\phi)$ in governing the RG observables.

The paper is organized as follows. In Sec.~\ref{sec:setup} we present the Einstein--scalar theory and define the superpotential-like function $W(\phi)$ from the equations of motion, showing that its near-boundary expansion reproduces the standard potential-superpotential relation to the order required for the ultraviolet analysis. In Sec.~\ref{sec:var} we analyze the variational principle under mixed boundary conditions, determine the asymptotic counterterm, and compute the renormalized Euclidean action and conserved energy. In Sec.~\ref{sec:rg} we study the holographic stress tensor together with the $c$-function and $\beta$-function associated with $W(\phi)$. In Sec.~\ref{sec:examples} we apply the formalism to exact asymptotically AdS black-hole solutions arising from consistent truncations of supergravity, with mixed boundary conditions that preserve conformal symmetry.

\section{Set-up}
\label{sec:setup}

We consider four-dimensional Einstein gravity minimally coupled to a self-interacting scalar field, described by the action
\begin{equation}
\label{eq:action}
I[g_{\mu\nu},\phi]=\frac{1}{2\kappa}\int_{\mathcal{M}}d^{4}x\sqrt{-g}\,\left[R-\frac{1}{2}(\partial\phi)^{2}-V(\phi)\right].
\end{equation}
We work in units where $c=1$, so that $\kappa=8\pi G$.

The Einstein equations and the scalar equation of motion are
\begin{equation}
\label{eq:eom}
E_{\mu\nu}:=G_{\mu\nu}-\frac{1}{2}\partial_{\mu}\phi\,\partial_{\nu}\phi+\frac{1}{2}g_{\mu\nu}\left[\frac{1}{2}(\partial\phi)^{2}+V(\phi)\right]=0,
\qquad
\nabla_{\mu}\nabla^{\mu}\phi-\frac{dV(\phi)}{d\phi}=0,
\end{equation}
respectively.

We restrict attention to static black-hole solutions of the form
\begin{equation}
\label{eq:metric}
ds^{2}=-N(r)dt^{2}+H(r)dr^{2}+S(r)\,d\Sigma_{k},
\qquad
d\Sigma_{k}\equiv \frac{dy^{2}}{1-ky^{2}}+(1-ky^{2})dz^{2},
\end{equation}
where $k=-1,0,1$ correspond to hyperbolic, planar, and spherical horizons, respectively.

The scalar potential is assumed to admit an expansion around $\phi=0$ of the form
\begin{equation}
\label{eq:Vexp}
V(\phi)=2\Lambda+\frac{1}{2}m^{2}\phi^{2}+\mathcal{O}(\phi^{4}),
\qquad
\Lambda\equiv -\frac{3}{L^{2}},
\end{equation}
and we focus on scalar fields with mass
\begin{equation}
\label{eq:mass}
m^{2}L^{2}=-2,
\end{equation}
which lies within the Breitenlohner--Freedman window and is therefore perturbatively stable \cite{Breitenlohner:1982bm,Breitenlohner:1982jf}. The scalar field admits the asymptotic expansion
\begin{equation}
\label{eq:falloff}
\phi(r)=\frac{A}{r}+\frac{B}{r^{2}}+\mathcal{O}(r^{-3}),
\end{equation}
where $r$ is the canonical radial coordinate, defined by $S(r)=r^{2}+\mathcal{O}(r^{-2})$, and the coefficients $A$ and $B$ correspond to the two normalizable fall-off modes. Since both modes are normalizable, asymptotically AdS boundary conditions alone do not select a unique variational problem, and one must specify an additional boundary condition of the form \cite{Witten:2001ua}
\begin{equation}
\label{eq:BA}
B=B(A).
\end{equation}
Different choices of $B(A)$ define different theories within the framework of designer gravity.

For later use in the variational problem and in holographic renormalization, it is convenient to introduce a superpotential-like function $W(\phi)$ whose near-boundary expansion captures the divergent structure of the on-shell action and encodes the mixed boundary condition for the scalar field. In practice, $W(\phi)$ also allows one to rewrite a subset of the second-order field equations in first-order form, thereby providing a solution-adapted Hamilton--Jacobi description along the radial direction. This function is not assumed to generate the bulk dynamics in the sense of supergravity; in particular, it does not satisfy Eq.~\eqref{eq:intro_sugra_potential} as an exact identity throughout the bulk.

Consider the combination of the Einstein equations $E^{t}_{t}-E^{r}_{r}$, which yields
\begin{equation}
\label{eq:EtEr}
(\phi')^{2}-\frac{(S')^{2}-2SS''}{S^{2}}-\frac{S'(NH)'}{SNH}=0,
\end{equation}
where a prime denotes differentiation with respect to $r$. Following \cite{Gursoy:2008za}, Eq.~\eqref{eq:EtEr} can be equivalently rewritten as the pair of first-order equations
\begin{equation}
\label{eq:Wdef}
W(\phi)=-\frac{2S'}{L(NHS)^{1/2}},
\qquad
\frac{dW(\phi)}{d\phi}=\phi' L\left(\frac{S}{NH}\right)^{1/2}.
\end{equation}
In the literature, such a quantity is often referred to as a thermal superpotential \cite{Anabalon:2015xvl}. We adopt the more neutral term \emph{superpotential-like function} to emphasize that $W(\phi)$ is an auxiliary, solution-dependent Hamilton--Jacobi-type object. It is defined directly from the equations of motion and will be used primarily through its near-boundary expansion to organize counterterms and implement mixed boundary conditions.

The remaining Einstein equation can be rearranged into the form
\begin{equation}
\label{eq:Veff}
V(\phi)=\frac{1}{2}\left[(\partial_{\phi}W)^{2}-\frac{3}{4}W^{2}\right]\frac{N L^{2}}{S}
-\frac{S'}{NH}\left(\frac{N}{S}\right)'+\frac{2k}{S},
\end{equation}
which closely resembles Eq.~\eqref{eq:intro_sugra_potential}, except for the explicit dependence on the metric functions. Using the standard AdS asymptotics \cite{Henneaux:2006hk}
\begin{equation}
\label{eq:NSasym}
N(r)=\frac{r^{2}}{L^{2}}+k-\frac{\mu}{r}+\mathcal{O}(r^{-2}),
\qquad
S(r)=r^{2}+\mathcal{O}(r^{-2}),
\end{equation}
where $\mu$ represents the leading deviation of the ground state,
and using the Einstein equations to determine
\begin{equation}
\label{eq:Hasym}
H(r)=\frac{L^{2}}{r^{2}}\left[1-\frac{A^{2}+4kL^{2}}{4r^{2}}-\frac{2AB-3L^{2}\mu}{3r^{3}}\right]+\mathcal{O}(r^{-6}),
\end{equation}
we straightforwardly find that Eq.~\eqref{eq:Veff} yields, at leading order,
\begin{equation}
\label{eq:Vasym}
V(\phi)=\frac{1}{2}\left[(\partial_{\phi}W)^{2}-\frac{3}{4}W^{2}\right]\left[1+\mathcal{O}(r^{-2})\right]+\mathcal{O}(r^{-2}),
\end{equation}
which asymptotically reduces to the standard Boucher--Townsend \cite{Townsend:1984iu,Boucher:1984yx}/Faulkner--Horowitz--Roberts form \cite{Faulkner:2010fh}, that is, to Eq. (\ref{eq:intro_sugra_potential}).

Equations~\eqref{eq:Vasym} and \eqref{eq:Wdef} determine the near-boundary expansion of $W(\phi)$. We therefore assume the power-series ansatz
\begin{equation}
\label{eq:ansatzW}
W(\phi)=\sum_{j\geq 0} a_{j}\phi^{j},
\end{equation}
The leading coefficient is fixed by Eq.~\eqref{eq:Wdef}. Using the asymptotic AdS behavior of the metric functions, one finds
\begin{equation}
\label{eq:a0}
W(\phi)=-\frac{2S'}{L(NHS)^{1/2}}
\qquad \Longrightarrow \qquad
a_{0}+\frac{4}{L}+\mathcal{O}(r^{-1})=0.
\end{equation}
Therefore $a_{0}=-4/L$. We then use Eq.~\eqref{eq:Vasym} together with the asymptotic expansion of the potential in Eq.~\eqref{eq:Vexp}. The right-hand side of Eq.~\eqref{eq:Vasym}, expanded up to quadratic order in $\phi$, takes the form
\begin{equation}
\label{eq:Vmatch}
V(\phi)=\frac{1}{2}a_{1}^{2}-\frac{3}{8}a_{0}^{2}
+\left(2a_{1}a_{2}-\frac{3}{4}a_{0}a_{1}\right)\phi
+\left(3a_{1}a_{3}+2a_{2}^{2}-\frac{3}{4}a_{0}a_{2}-\frac{3}{8}a_{1}^{2}\right)\phi^{2}
+\mathcal{O}(\phi^{3}).
\end{equation}
Matching this expression with the asymptotic expansion of the scalar potential in Eq.~\eqref{eq:Vasym} yields the algebraic system
\begin{equation}
\label{eq:algsys}
\frac{1}{2}a_{1}^{2}-\frac{3}{8}a_{0}^{2}=-\frac{6}{L^{2}},
\qquad
2a_{1}a_{2}-\frac{3}{4}a_{0}a_{1}=0,
\qquad
3a_{1}a_{3}+2a_{2}^{2}-\frac{3}{4}a_{0}a_{2}-\frac{3}{8}a_{1}^{2}=-\frac{1}{L^{2}}.
\end{equation}
Choosing the simplest branch compatible with the AdS critical point at $\phi=0$, namely
\begin{equation}
\label{eq:branch}
a_{0}=-\frac{4}{L},
\qquad
a_{1}=0,
\qquad
a_{2}=-\frac{1}{2L},
\end{equation}
we obtain the near-boundary expansion
\begin{equation}
\label{eq:WUV}
W(\phi)=-\frac{4}{L}-\frac{\phi^{2}}{2L}+a\phi^{3}+\mathcal{O}(\phi^{4}),
\end{equation}
where the cubic coefficient $a_{3}\equiv a$ remains undetermined at this stage.

\section{Variational principle and regularization}
\label{sec:var}

In Einstein--scalar theories with asymptotically AdS boundary conditions and mixed scalar boundary conditions \eqref{eq:BA}, a well-defined variational principle and a finite on-shell action require boundary terms beyond the Gibbons--Hawking contribution \cite{deBoer:1999tgo,Hertog:2004dr,Anabalon:2015xvl,deHaro:2000vlm}. These additional terms must simultaneously cancel ultraviolet divergences and ensure the consistency of the variational problem under variations compatible with \eqref{eq:BA}. A particularly convenient framework is to construct the scalar boundary term from the superpotential-like function $W(\phi)$. Since both the ultraviolet divergences and the variational consistency are controlled by the near-boundary region, only the asymptotic expansion of $W(\phi)$ around $\phi=0$ is relevant \cite{deHaro:2000vlm}.
In particular, it suffices that the scalar potential $V(\phi)$ admit
an asymptotic superpotential representation in the UV, as in (\ref{eq:Vasym}), so that $W(\phi)$ matches a genuine superpotential up to subleading corrections.

In what follows, we analyze the variational problem for the renormalized action in the presence of mixed boundary conditions for the scalar field \cite{Papadimitriou:2004ap,Hertog:2004dr,Anabalon:2015xvl,Batrachenko:2004fd}. We show that the coefficient a in (\ref{eq:WUV}) is not determined by the bulk equations alone, but is instead fixed by demanding a well-posed variational principle. We then evaluate the Euclidean on-shell action and verify the quantum–statistical relation for the class of designer-gravity theories under consideration.

\subsection{Variational principle under mixed boundary conditions}

We consider the total Euclidean action
\begin{equation}
\label{eq:IE}
I^{\text{E}}=
-\frac{1}{2\kappa}\int_{\mathcal{M}}d^{4}x\sqrt{g^{\text{E}}}\left[R-\frac{1}{2}(\partial\phi)^{2}-V(\phi)\right]
-\frac{1}{\kappa}\int_{\partial\mathcal{M}}d^{3}x\sqrt{h^{\text{E}}}\,K
+I_{W}^{\text{E}}+I_{\text{ct}}^{\text{E}},
\end{equation}
where $K$ is the trace of the extrinsic curvature, evaluated on the $r=\text{const}$ hypersurface with induced metric $ds^{2}_{\partial\mathcal{M}}=h_{ab}dx^{a}dx^{b}$. 
The term $I_{W}^{\text{E}}$ is the scalar boundary term constructed from the
superpotential-like function $W(\phi)$, whereas $I_{\text{ct}}^{\text{E}}$
denotes the purely gravitational counterterm required to render
the on-shell action finite. These terms are given by \cite{Batrachenko:2004fd, Anabalon:2025sqr}
\begin{equation}
\label{eq:IWct}
I_{W}^{\text{E}}=-\frac{1}{2\kappa}\int_{\partial\mathcal{M}}d^{3}x\sqrt{h^{\text{E}}}\,W(\phi),
\qquad
I_{\text{ct}}^{\text{E}}=\frac{L}{2\kappa}\int_{\partial\mathcal{M}}d^{3}x\sqrt{h^{\text{E}}}\,\mathcal{R}^{(3)},
\end{equation}
where $\mathcal{R}^{(3)}=2k/S(r)$ is the Ricci scalar of the hypersurface $r=\text{const.}$

We impose mixed boundary conditions for the scalar field, specified by the functional relation \eqref{eq:BA} between the leading and subleading modes in the asymptotic expansion \eqref{eq:falloff}. This relation removes the independence of the two asymptotic modes. Therefore, when varying the action at the boundary, $A$ and $B$ cannot be varied independently, but their variations satisfy $\delta B=\qty(dB/dA)\delta A$.

Using this relation, the on-shell variation of the Euclidean action \eqref{eq:IE} reduces to a boundary term proportional to $\delta A$,
\begin{equation}
\label{eq:dIE}
\delta I^{\text{E}}
=
\frac{1}{2\kappa}\int_{\partial\mathcal{M}}d^{3}x\sqrt{h^{E}}
\left[n^{\mu}\partial_{\mu}\phi\,\delta\phi-\delta W(\phi)\right]
=
-\frac{\beta\sigma_{k}}{2\kappa L^{2}}\left[\left(3a+\dv{a}{A}A\right)LA^{2}+B\right]\delta A+\mathcal{O}(r^{-1}),
\end{equation}
where $\sigma_{k}$ denotes the area of the unit transverse section, $\beta$ is the periodicity of Euclidean-time required for regularity, and $n^{\mu}$ is the outward-pointing unit normal to the hypersurface $r=\text{const}$.

Requiring the variational principle to be well posed, $\delta I^{E}=0$, therefore implies
\begin{equation}
\label{eq:aeq}
\frac{d~}{dA}(A^{3}a(A))+\frac{B(A)}{L}=0 ~\Rightarrow~ a(A)=-\frac{1}{LA^{3}}\int B\,dA.
\end{equation}
Consistency of the variational principle then requires the mixed boundary condition to be integrable, namely that $B(A)$ can be written as the derivative of a boundary functional. That is, there must exist a function $w(A)$ such that
\begin{equation}
\label{eq:wdef}
B=\frac{dw(A)}{dA}.
\end{equation}
Equation \eqref{eq:aeq} is then immediately integrated to give
\begin{equation}
\label{eq:asol}
a(A)=\frac{w_{0}-w(A)}{LA^{3}},
\end{equation}
where $w_{0}$ is an arbitrary constant. Substituting \eqref{eq:asol} into \eqref{eq:WUV}, we obtain
\begin{equation}
\label{eq:Wmixed}
W(\phi)=-\frac{4}{L}-\frac{\phi^{2}}{2L}+\frac{w_{0}-w(A)}{LA^{3}}\phi^{3}+\mathcal{O}(\phi^{4}).
\end{equation}
Thus, the function $w(A)$ enters explicitly in the asymptotic expansion of the superpotential-like function
$W(\phi)$, thereby encoding the mixed boundary condition directly in the asymptotic counterterm.
\subsection{Euclidean on-shell action}
Each contribution to the total action \eqref{eq:IE} can be evaluated explicitly on shell. The bulk term is obtained by integrating the Lagrangian density after using the Einstein equation $E^{t}_{~t}=0$ to eliminate the scalar potential. One finds
\begin{equation}
\label{eq:Ibulk}
I_{\mathrm{bulk}}^{\text{E}}
=
\frac{\beta\sigma_{k}}{2\kappa}
\left[\frac{S N'}{\sqrt{HN}}\right]_{r_{h}}^{r}
=
\frac{\beta\sigma_{k}}{2\kappa}
\left(
\frac{2r^{3}}{L^{2}}+\frac{A^{2}r}{4L^{2}}+\mu+\frac{2AB}{3L^{2}}
\right)
-\beta T \mathcal{S}_{\mathrm{BH}}+\mathcal{O}(r^{-1}),
\end{equation}
where $S_{BH}$ is the Bekenstein--Hawking entropy, and $\beta = T^{-1}$ is the Euclidean time period fixed by regularity at the horizon. Explicitly,
\begin{equation}
\label{eq:SBH}
\mathcal{S}_{\mathrm{BH}}=\frac{\mathcal{A}_h}{4G}=\frac{\sigma_{k}S(r_{h})}{4G},
\qquad
\beta=4\pi\,\frac{\sqrt{H(r_{h})N(r_{h})}}{N'(r_{h})}\equiv T^{-1}.
\end{equation}
The Gibbons--Hawking contribution is
\begin{equation}
\label{eq:IGH}
I_{\mathrm{GH}}^{\text{E}}
=
\frac{\beta\sigma_{k}}{2\kappa}
\left[
-\frac{6r^{3}}{L^{2}}-\left(4k+\frac{3A^{2}}{4L^{2}}\right)r+3\mu-\frac{2AB}{L^{2}}
\right]
+\mathcal{O}(r^{-1}),
\end{equation}
whereas the boundary term constructed from the superpotential-like function gives
\begin{equation}
\label{eq:IW}
I_{W}^{\text{E}}
=
\frac{\beta\sigma_{k}}{2\kappa}
\left[
\frac{4r^{3}}{L^{2}}+\left(2k+\frac{A^{2}}{2L^{2}}\right)r-2\mu+\frac{(B-LA^{2}a)A}{L^{2}}
\right]
+\mathcal{O}(r^{-1}),
\end{equation}
Finally, the purely geometric counterterm contributes
\begin{equation}
\label{eq:Ict}
I_{\mathrm{ct}}^{E}
=
\frac{\beta\sigma_{k}}{\kappa}\,k\,r+\mathcal{O}(r^{-1}).
\end{equation}
Adding all contributions and taking the limit $r\to\infty$, the cubic and linear divergences cancel, and the renormalized Euclidean on-shell action becomes
\begin{equation}
\label{eq:Ionshell}
I^{\text{E}}
=
\frac{\beta\sigma_{k}}{4\pi}
\left\{
\frac{\mu}{2G}
+\frac{1}{4GL^{2}}\left[w(A)-w_{0}-\frac{1}{3}AB\right]
\right\}
-\beta T \mathcal{S}_{\mathrm{BH}},
\end{equation}
where in the last step we used Eqs.~\eqref{eq:asol} and \eqref{eq:Wmixed}.

\subsection{Quasilocal formalism and conserved energy}

Although the variational problem and the on-shell action discussed above were formulated in Euclidean signature, the Brown–York stress tensor and the associated quasilocal charges are naturally defined in Lorentzian signature. In this subsection, we therefore work with the Lorentzian continuation of the renormalized action corresponding to (\ref{eq:IE}).

In the Brown--York quasilocal formalism \cite{Brown:1992br}, the total energy is defined as the conserved charge associated with the timelike Killing vector that generates time translations. For a timelike boundary foliated by spacelike two-surfaces $\Sigma$, the corresponding quasilocal energy is
\begin{equation}
\label{eq:BYE}
E=\int_{\Sigma} d^{2}y\sqrt{\sigma}\,u^{a}\tau_{ab}\xi^{b},
\end{equation}
where $\sigma$ denotes the determinant of the induced metric on $\Sigma$, $u^{a}$ is the future-directed unit normal to $\Sigma$ within the timelike boundary, $\xi^{a}=\delta^{a}_{t}$ is the timelike Killing vector generating time translations, and $\tau_{ab}$ is the Brown--York boundary stress tensor,
\begin{equation}
\label{eq:tau}
\tau^{ab}\equiv \frac{2}{\sqrt{-h}}\frac{\delta I}{\delta h_{ab}}.
\end{equation}
For the renormalized action corresponding to \eqref{eq:IE}, the boundary stress tensor takes the form
\begin{equation}
\label{eq:tautotal}
\tau_{ab}
=
-\frac{1}{8\pi G}
\left[
K_{ab}-h_{ab}K-LG_{ab}+\frac{W(\phi)}{2L}h_{ab}
\right].
\end{equation}
Here $K_{ab}$ denotes the extrinsic curvature of the boundary embedded in the bulk spacetime, $K=h^{ab}K_{ab}$ is its trace, and $G_{ab}$ is the Einstein tensor constructed from the induced metric $h_{ab}$.

Evaluating the Brown–York charge for the metric ansatz \eqref{eq:metric} and the counterterms introduced above, we find
\begin{equation}
\label{eq:energy}
E=
\frac{\sigma_{k}}{16\pi G}\lim_{r\to\infty}
\frac{\sqrt{N}}{\sqrt{H}}
\left[
(2kL-SW)\sqrt{H}-2S'
\right]
=
\frac{\sigma_{k}}{4\pi}
\left\{
\frac{\mu}{2G}
+\frac{1}{4GL^{2}}\left[w(A)-w_{0}-\frac{1}{3}AB\right]
\right\}.
\end{equation}
This coincides precisely with the energy inferred from the Euclidean on-shell action \eqref{eq:Ionshell}. Therefore, the quantum-statistical relation
\begin{equation}
\label{eq:qsr}
\beta^{-1}I^{\text{E}}=E-TS_{\mathrm{BH}}
\end{equation}
is satisfied. This agreement provides a nontrivial consistency check between the Euclidean and quasilocal formulations of the renormalized theory under mixed boundary conditions.

\section{Holographic RG data from the superpotential-like function}
\label{sec:rg}

Within the AdS/CFT correspondence, the radial direction of the bulk geometry provides a geometric realization of the renormalization group of the dual field theory \cite{deBoer:1999tgo,Skenderis:2002wp,deBoer:2000cz}. In the present setup, the renormalized Brown–York tensor yields the holographic stress tensor in the standard way \cite{Skenderis:2002wp,deHaro:2000vlm,Balasubramanian:1999re}, while the solution-dependent superpotential-like function $W(\phi)$ provides a natural framework for organizing additional RG quantities associated with a given black-hole branch. In particular, $W(\phi)$ allows one to define branch-dependent holographic $c$- and $\beta$-functions in non-extremal backgrounds \cite{Anabalon:2025sqr,Anabalon:2020qux}.
For $m^2L^2=-2$ in AdS$_4$, the dual scalar operator admits the two standard quantizations, with scaling dimensions $\Delta_-=1$ and $\Delta_+=2$. Depending on the choice of quantization, either $A$ or $B$ is interpreted as the source for the dual operator $\mathcal{O}$, while the remaining coefficient is proportional to its expectation value \cite{Papadimitriou:2004ap,Witten:2001ua}. In the dual description, mixed boundary conditions are naturally interpreted as multi-trace deformations of the field-theory action, schematically of the form
\begin{equation}
I_{\mathrm{CFT}} \;\to\; I_{\mathrm{CFT}} - \int d^3x\, w(\mathcal{O}) \, .
\end{equation}
Accordingly, the precise meaning of the deformation depends on the holographic quantization. In the parametrization adopted here, this information is encoded in the function $w(A)$, whose derivative determines the mixed boundary condition $B=B(A)$ and whose contribution appears through finite boundary terms in the renormalized thermodynamic potential, without modifying the bulk equations of motion \cite{Witten:2001ua,Compere:2008us}.

In this section, we first review the construction of the holographic stress tensor in the present setup, emphasizing the role of $W(\phi)$ in non-extremal geometries. We then use this framework to define
a holographic c-function and a holographic $\beta$-function, generalizing the domain-wall results of \cite{Freedman:1999gp} to spacetimes with horizons and mixed scalar boundary conditions, along the lines of \cite{Gursoy:2008za}.

\subsection{Holographic stress tensor}

The dual field theory is defined on the boundary metric obtained by removing the divergent conformal factor from the induced metric on the hypersurfaces $r=\mathrm{const}$. This gives
\begin{equation}
\label{eq:dualmetric}
ds_{\mathrm{dual}}^{2}
=
\lim_{r\to\infty}\frac{L^{2}}{r^{2}}\,ds^{2}\big|_{\partial\mathcal{M}}
=
-dt^{2}+\gamma_{ij}dx^{i}dx^{j}
=
-dt^{2}+L^{2}\left[\frac{dy^{2}}{1-ky^{2}}+(1-ky^{2})dz^{2}\right].
\end{equation}
The expectation value of the stress tensor in the dual theory is obtained from the renormalized Brown--York tensor by the standard holographic rescaling \cite{Balasubramanian:1999re,Myers:1999psa}:
\begin{equation}
\label{eq:holorescale}
\langle T_{ab}\rangle=\lim_{r\to\infty}\frac{r}{L}\tau_{ab}.
\end{equation}
For the static backgrounds considered here, the coefficients $A$ and $B$ in the scalar asymptotics are constants (\ref{eq:falloff}). Evaluating the limit in Eq.~\eqref{eq:holorescale} using the results of the previous section, we obtain
\begin{equation}
\label{eq:Ttt}
\langle T_{tt}\rangle=
\frac{1}{4\pi}
\left\{
\frac{\mu}{2G}
+\frac{1}{4GL^{2}}\left[w(A)-w_{0}-\frac{1}{3}AB(A)\right]
\right\},
\end{equation}
\begin{equation}
\label{eq:Tij}
\langle T_{ij}\rangle=
\frac{\gamma_{ij}}{8\pi}
\left\{
\frac{\mu}{2G}
-\frac{1}{2GL^{2}}\left[w(A)-w_{0}-\frac{1}{3}AB(A)\right]
\right\},
\end{equation}
Hence the trace of the holographic stress tensor is
\begin{equation}
\label{eq:trace}
\langle T^{a}{}_{a}\rangle\equiv -\langle T_{tt}\rangle+\gamma^{ij}\langle T_{ij}\rangle
=
-\frac{3}{16\pi GL^{4}}
\left[
w(A)-w_{0}-\frac{1}{3}AB(A)
\right].
\end{equation}
For homogeneous boundary data, Eq.~\eqref{eq:trace} is the holographic trace Ward identity in the presence of mixed scalar boundary conditions. It makes explicit how conformal symmetry is broken by the scalar deformation encoded in the function $w(A)$.

Conformal invariance is recovered when the trace vanishes. This requires
\begin{equation}
\label{eq:conformalcond}
w(A)-w_{0}-\frac{1}{3}AB(A)=0
\qquad \Longrightarrow \qquad
w(A)-w_{0}\propto A^{3}
\qquad \Longrightarrow \qquad
B(A)\propto A^{2}.
\end{equation}
In the $\Delta_{-}=1$ quantization, this corresponds to a marginal triple-trace deformation of the dual three-dimensional theory.

The same condition can also be recovered from asymptotic AdS symmetry. Requiring the mixed scalar boundary condition to be compatible with the asymptotic AdS symmetry algebra implies that the combined fall-offs of the metric and scalar field must be preserved by the asymptotic Killing vectors \cite{Henneaux:2006hk,Henneaux:1985tv,Anabalon:2014fla}. For an asymptotically AdS metric written as
\begin{equation}
g_{\mu\nu}=g^{\mathrm{AdS}}_{\mu\nu}+h_{\mu\nu},
\end{equation}
the asymptotic symmetry algebra is generated by vector fields $\xi^{\mu}$ such that
\begin{equation}
\mathcal{L}_{\xi}g_{\mu\nu}=O(h_{\mu\nu})
\qquad (r\to\infty).
\end{equation}
Solving this condition yields
\begin{equation}
\xi^{r}=r\,\eta(x^{m})+O(r^{-1}),
\qquad
\xi^{m}=O(1),
\end{equation}
where $m=(t,y,z)$ and $\eta$ and $\xi^{m}$ are arbitrary functions of the boundary coordinates.

Allowing now for generic boundary data,
\begin{equation}
\phi(r,x^{m})
=
\frac{A(x^{m})}{r}
+
\frac{B(x^{m})}{r^{2}}
+
O(r^{-3}),
\end{equation}
its transformation under asymptotic symmetries implies
\begin{equation}
A\,\eta(x^{m})=\xi^{m}\partial_{m}A,
\qquad
2B\,\eta(x^{m})=\xi^{m}\partial_{m}B.
\end{equation}
Using $B=B(A)$, these relations combine into
\begin{equation}
\left(
2B-A\frac{dB}{dA}
\right)\xi^{m}\partial_{m}A=0.
\end{equation}
Requiring the asymptotic symmetry to be preserved for generic boundary data $A(x^{m})$ then implies
\begin{equation}
2B-A\frac{dB}{dA}=0,
\end{equation}
which again gives
\begin{equation}
B\propto A^{2},
\end{equation}
in agreement with Eq.~\eqref{eq:conformalcond}.

\subsection{The holographic \texorpdfstring{$c$}{c}-function}

At zero temperature, Poincar\'e-invariant holographic renormalization-group flows in Einstein--scalar theories can often be formulated in terms of a true or fake superpotential, allowing the second-order field equations to be rewritten as a first-order flow system. In that setting, and under the appropriate energy conditions, one can construct a holographic $c$-function that is monotonic along the flow and reproduces the conformal data at AdS fixed points \cite{deBoer:1999tgo,Freedman:1999gp,Skenderis:1999mm}.

At finite temperature, however, the presence of a regular black-hole horizon generically obstructs a global superpotential description in terms of a single function characterizing an entire family of Lorentz-invariant vacuum flows. Nevertheless, for any given solution one may still reorganize the radial dynamics in terms of a solution-dependent superpotential-like function $W(\phi)$ together with an associated blackening function. In this sense, the resulting holographic $c$-function should be understood as a branch-dependent geometric diagnostic attached to a given background, rather than as a universal quantity defined on the full space of flows.

To make this precise, we consider the generalized domain-wall gauge ansatz
\begin{equation}
\label{eq:DWmetric}
ds^{2}=e^{2A(u)}\left[-e^{g(u)}dt^{2}+L^{2}d\Sigma_{k}\right]+e^{-g(u)}du^{2},
\end{equation}
where $d\Sigma_k$ is the line element on a unit two-dimensional maximally symmetric space of curvature $k=0,\pm 1$. The function $A(u)$ plays the role of the logarithm of the field-theory energy scale, while $g(u)$ measures the departure from the zero-temperature domain-wall geometry.

From the Einstein equations, the combination $E^{t}{}_{t}-E^{u}{}_{u}=0$ yields
\begin{equation}
\label{eq:Wdef1}
4A''(u)+\bigl[\phi'(u)\bigr]^2=0,
\end{equation}
where primes denotes now derivative with respect to $u$. Therefore
\begin{equation}
A''(u)\leq 0.
\end{equation}
Choosing the radial orientation so that $A'(u)>0$, it follows that
\begin{equation}
\label{eq:monotonic}
\frac{d}{du}\!\left[\frac{1}{(A')^2}\right]
=
-\frac{2A''(u)}{[A'(u)]^3}
\geq 0.
\end{equation}
This motivates the definition
\begin{equation}
\label{eq:cdef}
c(u)=\frac{c_0}{L^2\,[A'(u)]^2},
\end{equation}
where $c_0$ is a dimensionless normalization constant fixed by matching to the ultraviolet CFT data. At an asymptotically AdS fixed point, the scalar approaches a constant and
\begin{equation}
\label{eq:AdSfixed}
A(u)=\frac{u}{L}
\qquad\Longrightarrow\qquad
A'(u)=\frac{1}{L},
\end{equation}
so that
\begin{equation}
c(u)\to c_0.
\end{equation}

Using the first-order relations compatible with Eq.~\eqref{eq:Wdef1},
\begin{equation}
\label{eq:firstorderDW}
4A''(u)+\bigl[\phi'(u)\bigr]^2=0~~\Rightarrow~~
\phi'(u)=\frac{dW(\phi)}{d\phi},
\qquad
A'(u)=-\frac{1}{4}W(\phi),
\end{equation}
the holographic $c$-function can be written directly in terms of the superpotential-like function as
\begin{equation}
c(\phi)=\frac{16\,c_0}{L^2\,W(\phi)^2}.
\end{equation}
Using the ultraviolet expansion of $W(\phi)$ obtained in Eq.~\eqref{eq:Wmixed}, one finds
\begin{equation}
c(\phi)=c_0+O(\phi^2),
\end{equation}
as expected near the AdS fixed point.

Although the definition of the $c$-function is most naturally formulated in domain-wall coordinates, the asymptotic expansion \eqref{eq:Wmixed} was extracted from the ultraviolet region of the finite-temperature black-hole solution. This is consistent because, near the AdS boundary, black-hole geometries are locally equivalent to domain-wall spacetimes and thermal effects are subleading. Accordingly, the same ultraviolet expansion of $W(\phi)$ can be used to characterize the near-boundary RG behavior of the non-extremal branch.

\subsection{The holographic \texorpdfstring{$\beta$}{beta}-function}

In holographic duality, renormalization-group flows in the boundary theory are geometrized as radial evolution in the bulk spacetime \cite{deBoer:1999tgo,Freedman:1999gp}. For a given choice of quantization, the scalar profile determines the running of the coupling associated with the dual operator. In many holographic constructions, it is convenient to identify the running coupling as
\begin{equation}
\lambda \equiv e^{\phi}.
\end{equation}
The holographic $\beta$-function is then defined by
\begin{equation}
\beta_{\lambda}(\lambda)\equiv \frac{d\lambda}{d\ln E},
\end{equation}
where $E$ denotes the energy scale of the dual field theory.

To evaluate this quantity, we use the generalized domain-wall gauge introduced in the previous subsection, namely Eq~\eqref{eq:DWmetric}.
In this parametrization, the function $A(u)$ controls the overall conformal factor of the induced metric and therefore sets the field-theory energy scale, $E\propto e^{A(u)}$, while $g(u)$ encodes the non-extremality of the background. It follows that
\begin{equation}
\beta_{\lambda}(\lambda)
=
\frac{d\lambda}{dA}
=
\left(\frac{dA}{du}\right)^{-1}\frac{d\lambda}{du}.
\end{equation}
Using $\lambda=e^{\phi}$, this becomes
\begin{equation}
\beta_{\lambda}(\lambda)
=
\left(\frac{dA}{du}\right)^{-1}\frac{d\phi}{du}\,e^{\phi}.
\end{equation}

Employing the first-order relations in Eq.~ \eqref{eq:firstorderDW},
\begin{equation}
\phi'(u)=\frac{dW(\phi)}{d\phi},
\qquad
A'(u)=-\frac{1}{4}W(\phi),
\end{equation}
we obtain
\begin{equation}
\beta_{\lambda}(\lambda)\equiv \left.\frac{d\phi}{d\ln E}\right|_{\phi=\ln\lambda}
=
-\frac{4}{W(\phi)}\frac{dW(\phi)}{d\phi}\,\lambda.
\end{equation}

Near an asymptotically AdS fixed point, the ultraviolet behavior of the $\beta$-function is determined by the asymptotic expansion of $W(\phi)$ given in Eq.~\eqref{eq:Wmixed}, which reproduces the expected perturbative structure. At finite temperature, the function $W(\phi)$ is solution dependent, and therefore so is the corresponding holographic $\beta$-function. Nevertheless, thermal effects do not modify the leading ultraviolet behavior, since the near-boundary expansion of $W(\phi)$ continues to control the asymptotic flow.

\section{Exact solutions: concrete cases}
\label{sec:examples}

We now illustrate the general formalism developed above in two explicit classes of asymptotically AdS black-hole solutions arising from consistent truncations of supergravity \cite{Anabalon:2020qux,Faedo:2015jqa,Faedo:2017veq}. These examples provide nontrivial checks of the construction presented in Secs.~2--4, since they allow the renormalized thermodynamic quantities and the holographic RG observables associated with the superpotential-like function $W(\phi)$ to be computed explicitly.

The two cases highlight complementary aspects of the formalism. In the first, $W(\phi)$ is reconstructed directly from the black-hole solution. In the second, the scalar potential also admits a supergravity superpotential $W_{\mathrm{SUGRA}}$, which allows for a direct comparison with the solution-dependent function extracted from the geometry. This comparison makes explicit why, for the non-extremal branch, the relevant counterterm structure and RG observables are governed by $W(\phi)$ rather than by $W_{\mathrm{SUGRA}}$.

\subsection{Hairy self-interacting black hole solutions}

As a first explicit test of the formalism, we consider the family of exact asymptotically AdS hairy black-hole solutions studied in \cite{Anabalon:2017yhv,Anabalon:2020qux}. These solutions arise in four-dimensional Einstein--scalar gravity with a nontrivial self-interacting dilaton potential and obey mixed boundary conditions for the scalar field. They have been extensively analyzed in the context of designer gravity, in particular with regard to their thermodynamic properties and phase structure \cite{Anabalon:2015xvl,Canfora:2021nca}.

The dilaton potential is
\begin{align}
\label{eq:Vhair}
V(\phi)
&=
2\alpha
\left[
\frac{\nu-1}{\nu+2}\sinh\left((\nu+1)l_{\nu}\phi\right)
-\frac{\nu+1}{\nu-2}\sinh\left((\nu-1)l_{\nu}\phi\right)
+\frac{4\nu^{2}-4}{\nu^{2}-4}\sinh(l_{\nu}\phi)
\right]
\nonumber\\
&\quad
-\frac{\nu^{2}-4}{\nu^{2}L^{2}}
\left[
\frac{\nu-1}{\nu+2}e^{-(\nu+1)l_{\nu}\phi}
+\frac{\nu+1}{\nu-2}e^{(\nu-1)l_{\nu}\phi}
+\frac{4\nu^{2}-4}{\nu^{2}-4}e^{-l_{\nu}\phi}
\right],
\end{align}
where $l_{\nu}\equiv(\nu^{2}-1)^{-1/2}$, $L$ is the AdS radius, $\alpha$ is a dimensionful constant of the theory, and $\nu$ is a dimensionless hair parameter. Near the AdS vacuum, the potential behaves as
\begin{equation}
\label{eq:Vhairas}
V(\phi)=-\frac{6}{L^{2}}-\frac{\phi^{2}}{L^{2}}-\frac{\nu^{2}-3}{12(\nu^{2}-1)L^{2}}\phi^{4}+\mathcal{O}(\phi^{5}),
\end{equation}
so the scalar has conformal mass $m^{2}=-2L^{-2}$.

The exact solution can be written as
\begin{equation}
\label{eq:hairsol}
ds^{2}=\Omega(x)\left[-f(x)dt^{2}+\frac{\eta^{2}dx^{2}}{f(x)}+d\Sigma_{k}^{2}\right],
\qquad
\phi(x)=l_{\nu}^{-1}\ln x,
\end{equation}
with
\begin{equation}
\label{eq:hairOmega}
\Omega(x)=\frac{\nu^{2}x^{\nu-1}}{\eta^{2}(x^{\nu}-1)^{2}},
\end{equation}
and
\begin{equation}
\label{eq:hairf}
f(x)=\frac{1}{L^{2}}+\frac{\eta^{2}(x^{\nu}-1)^{2}}{x^{\nu-2}\nu^{2}}k
+\frac{\alpha}{\nu^{2}}
\left[
\frac{\nu^{2}}{\nu^{2}-4}
-x^{2}
+\frac{x^{\nu+2}}{\nu+2}
-\frac{x^{2-\nu}}{\nu-2}
\right].
\end{equation}
In this coordinate system, the boundary is located at $x\to 1$, and there are two independent domains: $0\leq x<1$ and $0<x\leq\infty$.

To match this solution with the general ansatz in Eq.~\eqref{eq:metric}, we identify
\begin{equation}
\label{eq:hairID}
N=\Omega f,
\qquad
H=\frac{\eta^{2}\Omega}{f},
\qquad
S=\Omega.
\end{equation}
The superpotential-like function is then reconstructed directly from the geometry through Eq.~\eqref{eq:Wdef}, which gives
\begin{equation}
\label{eq:Whairdef}
W(\phi)=-\frac{2\Omega'}{\eta L\,\Omega^{3/2}},
\qquad
\partial_{\phi}W(\phi)=\phi' \eta L\,\Omega^{1/2}.
\end{equation}
where prime denotes derivative with respect to $x$.
Substituting the explicit form of $\Omega(x)$ and using $x=e^{l_{\nu}\phi}$, we obtain
\begin{equation}
\label{eq:Whair}
W(\phi)=
-\frac{2}{L\nu}
\left[
(\nu+1)e^{\frac{\nu-1}{2}l_{\nu}\phi}
+
(\nu-1)e^{-\frac{\nu+1}{2}l_{\nu}\phi}
\right].
\end{equation}
At the boundary where $\phi\approx 0$, the expansion becomes
\begin{equation}
\label{eq:WhairUV}
W(\phi)=
-\frac{1}{L}\left(
4+\frac{\phi^{2}}{2}+\frac{l_{\nu}}{6}\phi^{3}
\right)
+\mathcal{O}(\phi^{4}),
\end{equation}
in agreement with the general asymptotic form in Eq.~\eqref{eq:Wmixed}.

Using the near-boundary change of coordinates $r^{2}=\Omega(x)$, the scalar field admits the asymptotic expansion
\begin{equation}
\label{eq:hairfalloff}
\phi(r)=\frac{A}{r}+\frac{B}{r^{2}}+\mathcal{O}(r^{-3}),
\qquad
A=-\frac{1}{l_{\nu}\eta},
\qquad
B=-\frac{1}{2l_{\nu}\eta^{2}}
=-\frac{l_{\nu}}{2}A^{2}.
\end{equation}
Therefore,
\begin{equation}
B\propto A^{2},
\end{equation}
so the mixed boundary condition preserves conformal symmetry. Equivalently,
\begin{equation}
\label{eq:hairw}
B=\frac{dw}{dA}
\qquad \Longrightarrow \qquad
w(A)=w_{0}+\frac{1}{3}AB.
\end{equation}

The thermodynamic quantities are
\begin{equation}
\label{eq:hairthermo}
E=\frac{\sigma_{k}\mu}{8\pi G}=\frac{\sigma_{k}(\alpha+3k\eta^{2})}{24\pi G\eta^{3}},
\end{equation}
\begin{equation}
\label{eq:hairTS}
T=\frac{f'(x_{h})}{4\pi\eta},
\qquad
S_{\mathrm{BH}}=\frac{\sigma_{k}\nu^{2}x_{h}^{\nu-1}}{4G\eta^{2}(x_{h}^{\nu}-1)^{2}},
\end{equation}
and the free energy is determined by the quantum-statistical relation,
\begin{equation}
\label{eq:hairF}
F=E-TS_{\mathrm{BH}}.
\end{equation}
The holographic stress tensor is traceless and takes the form
\begin{equation}
\label{eq:hairstress}
\langle T_{tt}\rangle=\frac{\mu}{8\pi G},
\qquad
\langle T_{ij}\rangle=\frac{\mu}{16\pi G}\gamma_{ij},
\qquad
\langle T^{a}{}_{a}\rangle=0
\end{equation}
as expected for conformal boundary conditions.

Finally, the corresponding holographic $c$-function and $\beta$-function are
\begin{equation}
\label{eq:hairc}
c(\phi)=\frac{4\nu^{2}e^{(\nu+1)l_{\nu}\phi}c_{0}}{\left[(\nu+1)e^{\nu l_{\nu}\phi}+\nu-1\right]^{2}}
=
\left(1-\frac{1}{4}\phi^{2}+\frac{1}{12}l_{\nu}\phi^{3}\right)c_{0}+\mathcal{O}(\phi^{4}),
\end{equation}
and, introducing $\lambda\equiv e^{\phi}$,
\begin{equation}
\label{eq:hairbeta}
\beta_{\lambda}(\lambda)
=
-\frac{4}{W(\phi)}\frac{dW(\phi)}{d\phi}\,\lambda
=
-\frac{1}{2l_{\nu}}
\frac{\left(\lambda^{\nu l_{\nu}}-1\right)\lambda}{(\nu+1)\lambda^{\nu l_{\nu}}+\nu-1}.
\end{equation}
Near the AdS fixed point one has $\phi \to 0$, and hence $\lambda=e^{\phi}\to 1$. It is therefore natural to expand the holographic $\beta$-function around
$\ln\lambda=\phi=0$. Since the flow equations are formulated in terms of the scalar field, the ultraviolet behavior is most conveniently written as an expansion in $\ln\lambda$:
\begin{equation}
\beta_{\lambda}(\lambda)
=
-\frac{1}{4}\ln\lambda
+
\frac{1}{8}(l_{\nu}-2)(\ln\lambda)^2
+
O\!\left((\ln\lambda)^3\right).
\end{equation}
This provides an explicit realization of the general construction developed in secs.~3 and 4.

\subsection{\texorpdfstring{$\mathcal{N}=2$}{N=2} gauged supergravity}

As a second example, we consider hairy black holes in four-dimensional $\mathcal{N}=2$ gauged supergravity \cite{Faedo:2015jqa,Faedo:2017veq}. In this family, the AdS$_4$ radius $L$ is fixed by the gauging of the underlying theory, while the parameter $n$ labels the one-parameter family of models within the single-scalar truncation. This example is particularly useful because, in contrast with the previous case, the scalar potential admits a genuine supergravity superpotential. It therefore provides a direct setting in which to compare the supergravity quantity $W_{\mathrm{SUGRA}}$ with the solution-dependent superpotential-like function $W(\phi)$ reconstructed from the black-hole background itself.

The scalar potential is
\begin{equation}
\label{eq:VsugraN2}
V(\phi)
=
\frac{n(1-2n)}{L^{2}}e^{\frac{n-2}{\sqrt{2}}\ell_{n}\phi}
+\frac{4n(n-2)}{L^{2}}e^{\frac{n-1}{\sqrt{2}}\ell_{n}\phi}
-\frac{(3-2n)(2-n)}{L^{2}}e^{\frac{n}{\sqrt{2}}\ell_{n}\phi},
\end{equation}
where
\begin{equation}
\label{eq:lambdan}
\ell_{n}\equiv \sqrt{\frac{2}{n(2-n)}},
\qquad
0<n<2.
\end{equation}
For this theory, one may introduce a superpotential satisfying the standard relation ~\eqref{eq:intro_sugra_potential},
\begin{equation}
\label{eq:WSUGRA}
W_{\text{SUGRA}}(\phi)=
\frac{2}{L}
\left[
n\,e^{-\sqrt{\frac{2-n}{n}}\frac{\phi}{2}}
+
(2-n)e^{\sqrt{\frac{n}{2-n}}\frac{\phi}{2}}
\right],
\end{equation}
with
\begin{equation}
\label{eq:VfromWS}
V(\phi)
=
\frac{1}{2}
\left[
\left(\frac{dW_{\mathrm{SUGRA}}}{d\phi}\right)^{2}
-
\frac{3}{4}W_{\mathrm{SUGRA}}^{2}
\right].
\end{equation}

The exact black-hole solution is
\begin{equation}
\label{eq:N2metric}
ds^{2}
=
-\frac{f(\rho)}{f_{1}^{n}(\rho)f_{2}^{2-n}(\rho)}\,d\tilde{t}^{2}
+
f_{1}^{n}(\rho)f_{2}^{2-n}(\rho)\left(\frac{d\rho^{2}}{f(\rho)}+d\Sigma_{k}\right),
\end{equation}
where
\begin{equation}
\label{eq:N2f12}
f_{1}(\rho)=\frac{n}{\sqrt{2}}\left(\rho+\frac{2\beta}{n}\right),
\qquad
f_{2}(\rho)=\frac{n}{\sqrt{2}}\left(\rho-\frac{2\beta}{2-n}\right),
\end{equation}
\begin{equation}
\label{eq:N2f}
f(\rho)=\frac{f_{1}(\rho)f_{2}(\rho)}{L^{2}}
\left[
\rho^{2}
-\frac{4(1-n)}{n(2-n)}\beta\rho
+4\beta^{2}\frac{5n^{2}-10n+4}{n^{2}(2-n)^{2}}
+\frac{2kL^{2}}{n^{2}}
\right],
\end{equation}
with scalar profile
\begin{equation}
\label{eq:N2phi}
\phi(\rho)=\frac{\sqrt{2}}{\ell_{n}}\ln\left(\frac{f_{2}}{f_{1}}\right).
\end{equation}
Here $\beta$ is the integration constant related to the energy of the system.

To match this background with the canonical ansatz in Eq.~\eqref{eq:metric}, we introduce the radial coordinate
\begin{equation}
\label{eq:N2coord}
r^{2}=f_{1}^{n}(\rho)f_{2}^{2-n}(\rho),
\end{equation}
together with the identifications
\begin{equation}
\label{eq:N2ID}
N(r)S(r)=f\big(\rho(r)\big)\left(\frac{d\tilde{t}}{dt}\right)^{2},
\qquad
\sqrt{N(r)H(r)}=\frac{d\tilde{t}}{dt}\frac{d\rho(r)}{dr},
\qquad
\tilde{t}=\frac{n}{\sqrt{2}}\,t.
\end{equation}
Solving the change of coordinates in the asymptotic region gives
\begin{equation}
\label{eq:N2rho}
\rho(r)=
\frac{\sqrt{2}}{n}r
+\frac{\sqrt{2}\beta^{2}}{(2-n)r}
-\frac{8(1-n)\beta^{3}}{3(2-n)^{2}r^{2}}
+\mathcal{O}(r^{-3}).
\end{equation}
Using this result, the scalar field admits the asymptotic expansion
\begin{equation}
\label{eq:N2falloff}
\phi(r)=\frac{A}{r}+\frac{B}{r^{2}}+\mathcal{O}(r^{-3}),
\qquad
A=-\sqrt{\frac{2}{2-n}}\,\sqrt{n}\,\beta,
\qquad
B=\frac{4\sqrt{n}(1-n)}{(2-n)^{3/2}}\,\beta^{2}.
\end{equation}
Therefore the solution satisfies the conformal mixed boundary condition
\begin{equation}
\label{eq:N2w}
B(A)=\frac{1}{4}\sqrt{2}\,\ell_{n}\sqrt{2-n}\,A^{2}
\qquad \Longrightarrow \qquad
w(A)=w_{0}+\frac{1}{3}AB.
\end{equation}

Adapting the general definition \eqref{eq:Wdef} to the present solution through Eq.~\eqref{eq:N2ID}, we obtain
\begin{equation}
\label{eq:N2Wdef}
W=-\frac{4}{L}\frac{dS^{1/2}}{d\rho}\frac{d\rho}{dr},
\qquad
S=f_{1}^{n}f_{2}^{2-n}.
\end{equation}
Expressed as a function of $\phi$, this yields
\begin{equation}
\label{eq:N2W}
W(\phi)=
-\frac{2}{L}
\left[
n\,e^{\sqrt{\frac{2-n}{n}}\frac{\phi}{2}}
+
(2-n)e^{-\sqrt{\frac{n}{2-n}}\frac{\phi}{2}}
\right].
\end{equation}
The two functions are related by
\begin{equation}
\label{eq:relationWS}
W_{\mathrm{SUGRA}}(\phi)=-W(-\phi).
\end{equation}
Although $W_{\mathrm{SUGRA}}(\phi)$ correctly reproduces the scalar potential, it does not replace the solution-adapted function $W(\phi)$ in the non-extremal black-hole branch. The reason is that the two functions have different near-boundary expansions and therefore do not generate the same scalar counterterm structure under mixed boundary conditions. Consequently,
\begin{equation}
\label{eq:betaineqN2}
-\frac{4}{W(\phi)}\frac{dW(\phi)}{d\phi}e^{\phi}
\neq
-\frac{4}{W_{\mathrm{SUGRA}}(\phi)}\frac{dW_{\mathrm{SUGRA}}(\phi)}{d\phi}e^{\phi}.
\end{equation}
This makes explicit that the RG observables of the non-extremal branch are governed by the function reconstructed from the geometry, rather than by the supergravity superpotential defined solely from the scalar potential.

Introducing the running coupling $\lambda=e^{\phi}$, the exact holographic $\beta$-function is
\begin{equation}
\label{eq:N2beta}
\beta_{\lambda}(\lambda)
=
-\frac{
2\sqrt{n(2-n)}
\left[
\lambda-\lambda^{\,1-\frac{1}{\sqrt{n(2-n)}}}
\right]
}{
n+(2-n)\lambda^{-\frac{1}{\sqrt{n(2-n)}}}
}
\end{equation}
Expanding around the AdS fixed point, where $\lambda \to 1$, one finds
\begin{equation}
\beta_{\lambda}(\lambda)
=
-(\lambda-1)
-\frac{1}{2}
\left(
1-\frac{n-1}{\sqrt{n(2-n)}}
\right)
(\lambda-1)^2
+O\!\left[(\lambda-1)^3\right].
\end{equation}
The corresponding holographic $c$-function is
\begin{equation}
c(\phi)
=
\frac{16c_{0}}{L^{2}W(\phi)^{2}}
=
c_{0}-\frac{c_{0}}{4}\phi^{2}+O(\phi^{3}),
\end{equation}
in agreement with the general ultraviolet behavior derived in Sec.~\ref{sec:rg}.

This example therefore provides the cleanest illustration of the conceptual distinction between $W_{\mathrm{SUGRA}}$ and the solution-dependent function $W(\phi)$: the former encodes the scalar potential, whereas the latter governs the counterterms and the RG observables relevant for the non-extremal black-hole background.

\section{Conclusions}

In this work we have revisited holographic renormalization and the variational problem for four-dimensional Einstein gravity coupled to a self-interacting scalar field in asymptotically AdS spacetimes with mixed (designer) boundary conditions. The central object throughout the analysis was the solution-dependent superpotential-like function $W(\phi)$, defined directly from the equations of motion and motivated by the Hamilton--Jacobi framework. The main result we would like to highlight is that, for the class of static solutions considered here, the near-boundary behavior of $W(\phi)$ is sufficient to control both the scalar counterterm required by holographic renormalization and the RG quantities associated with the corresponding black-hole branch.

Focusing on the case $m^{2}L^{2}=-2$, we showed that the asymptotic expansion of $W(\phi)$
is not fixed completely by the bulk equations alone. Rather, once integrable mixed boundary conditions $B=B(A)$ are imposed, the requirement of a well-posed variational principle determines the cubic coefficient in terms of the boundary deformation $w(A)$,
\begin{equation}
W(\phi)
=
-\frac{4}{L}
-\frac{\phi^{2}}{2L}
+
\frac{w_{0}-w(A)}{LA^{3}}\phi^{3}
+
O(\phi^{4})
\end{equation}
In this way, the mixed boundary condition is incorporated directly into the scalar counterterm, so that no additional scalar boundary term is needed beyond the standard geometric contributions.

Using this asymptotic form of $W(\phi)$, we obtained a finite Euclidean on-shell action and a finite quasilocal energy under mixed boundary conditions. We also showed that the two are related by the expected quantum-statistical relation, providing a nontrivial consistency check between the Euclidean and Brown--York formulations of the renormalized theory. On the holographic side, the same framework yields the renormalized stress tensor and its trace Ward identity, making explicit how the boundary deformation encoded in $w(A)$ modifies the dual theory.

A second main outcome of the paper is that the same superpotential-like function $W(\phi)$ also organizes the holographic RG data of non-extremal backgrounds. In particular, it determines a natural holographic $\beta$-function and a holographic $c$-function whose ultraviolet behavior is governed by the near-boundary expansion of the solution itself. In this sense, the relevant RG quantities at finite temperature are not controlled by a superpotential defined solely from the scalar potential, but by the function reconstructed from the black-hole geometry. They should therefore be understood as branch-dependent geometric diagnostics associated with a given background.

We illustrated this structure in two explicit classes of asymptotically AdS hairy black-hole solutions arising from consistent truncations of supergravity. In the first example, the formalism reproduces the expected thermodynamic and holographic quantities directly from the superpotential-like function extracted from the geometry. In the second, the comparison with a genuine supergravity superpotential $W_{\mathrm{SUGRA}}$ makes the conceptual distinction fully explicit: although $W_{\mathrm{SUGRA}}$ correctly reproduces the scalar potential, it does not in general determine the scalar counterterm structure or the RG observables of the non-extremal branch. These are instead governed by the solution-dependent function $W(\phi)$.

Several extensions deserve further study. A natural next step is to generalize the analysis to other scalar masses within the Breitenlohner--Freedman window and to more general classes of mixed boundary conditions, in order to determine how broadly the variational-principle criterion continues to fix the asymptotic counterterms. It would also be interesting to extend the construction to electrically charged hairy black holes, where the coupling between the scalar and gauge sectors is expected to lead to a richer thermodynamic and RG structure.

\section*{\normalsize Acknowledgments}
\vspace{-5pt}
The work of DC was supported by Pontificia Universidad Cat\'olica del Per\'u. The work of RR was partially supported by FONDECYT Postdoc Grant 3220663.

\bibliographystyle{JHEP}
\bibliography{bibliographyBH}

@article{deBoer:1999tgo,
  author        = "de Boer, Jan and Verlinde, Erik P. and Verlinde, Herman L.",
  title         = "{On the Holographic Renormalization Group}",
  eprint        = "hep-th/9912012",
  archivePrefix = "arXiv",
  journal       = "JHEP",
  volume        = "0008",
  pages         = "003",
  year          = "2000"
}

@article{Bianchi:2001kw,
    author = "Bianchi, Massimo and Freedman, Daniel Z. and Skenderis, Kostas",
    title = "{Holographic renormalization}",
    eprint = "hep-th/0112119",
    archivePrefix = "arXiv",
    reportNumber = "MIT-CTP-3166, PUTP-1999, DAMTP-2001-63, ROM2F-2001-30",
    doi = "10.1016/S0550-3213(02)00179-7",
    journal = "Nucl. Phys. B",
    volume = "631",
    pages = "159--194",
    year = "2002"
}

@article{Skenderis:2002wp,
    author = "Skenderis, Kostas",
    editor = "de Wit, B. and Vandoren, S.",
    title = "{Lecture notes on holographic renormalization}",
    eprint = "hep-th/0209067",
    archivePrefix = "arXiv",
    reportNumber = "PUTP-2047",
    doi = "10.1088/0264-9381/19/22/306",
    journal = "Class. Quant. Grav.",
    volume = "19",
    pages = "5849--5876",
    year = "2002"
}

@article{deBoer:2000cz,
    author = "de Boer, Jan",
    editor = "Bergshoeff, E. A. and Ceresole, Anna and Kounnas, C. and Lust, D. and Sevrin, A.",
    title = "{The Holographic renormalization group}",
    eprint = "hep-th/0101026",
    archivePrefix = "arXiv",
    doi = "10.1002/1521-3978(200105)49:4/6<339::AID-PROP339>3.0.CO;2-A",
    journal = "Fortsch. Phys.",
    volume = "49",
    pages = "339--358",
    year = "2001"
}

@article{Papadimitriou:2004ap,
    author = "Papadimitriou, Ioannis and Skenderis, Kostas",
    editor = "Biquard, O.",
    title = "{AdS / CFT correspondence and geometry}",
    eprint = "hep-th/0404176",
    archivePrefix = "arXiv",
    reportNumber = "ITFA-2004-17",
    doi = "10.4171/013-1/4",
    journal = "IRMA Lect. Math. Theor. Phys.",
    volume = "8",
    pages = "73--101",
    year = "2005"
}

@article{Maldacena:1997re,
    author = "Maldacena, Juan Martin",
    title = "{The Large N limit of superconformal field theories and supergravity}",
    eprint = "hep-th/9711200",
    archivePrefix = "arXiv",
    reportNumber = "HUTP-97-A097, HUTP-98-A097",
    doi = "10.4310/ATMP.1998.v2.n2.a1",
    journal = "Adv. Theor. Math. Phys.",
    volume = "2",
    pages = "231--252",
    year = "1998"
}

@article{Witten:1998qj,
    author = "Witten, Edward",
    title = "{Anti-de Sitter space and holography}",
    eprint = "hep-th/9802150",
    archivePrefix = "arXiv",
    reportNumber = "IASSNS-HEP-98-15",
    doi = "10.4310/ATMP.1998.v2.n2.a2",
    journal = "Adv. Theor. Math. Phys.",
    volume = "2",
    pages = "253--291",
    year = "1998"
}

@article{Gubser:1998bc,
    author = "Gubser, S. S. and Klebanov, Igor R. and Polyakov, Alexander M.",
    title = "{Gauge theory correlators from noncritical string theory}",
    eprint = "hep-th/9802109",
    archivePrefix = "arXiv",
    reportNumber = "PUPT-1767",
    doi = "10.1016/S0370-2693(98)00377-3",
    journal = "Phys. Lett. B",
    volume = "428",
    pages = "105--114",
    year = "1998"
}

@article{Klebanov:1999tb,
    author = "Klebanov, Igor R. and Witten, Edward",
    title = "{AdS / CFT correspondence and symmetry breaking}",
    eprint = "hep-th/9905104",
    archivePrefix = "arXiv",
    reportNumber = "PUPT-1863, IASSNS-HEP-99-49",
    doi = "10.1016/S0550-3213(99)00387-9",
    journal = "Nucl. Phys. B",
    volume = "556",
    pages = "89--114",
    year = "1999"
}

@article{Hertog:2004dr,
    author = "Hertog, Thomas and Maeda, Kengo",
    title = "{Black holes with scalar hair and asymptotics in N = 8 supergravity}",
    eprint = "hep-th/0404261",
    archivePrefix = "arXiv",
    doi = "10.1088/1126-6708/2004/07/051",
    journal = "JHEP",
    volume = "07",
    pages = "051",
    year = "2004"
}

@article{Witten:2001ua,
    author = "Witten, Edward",
    title = "{Multitrace operators, boundary conditions, and AdS / CFT correspondence}",
    eprint = "hep-th/0112258",
    archivePrefix = "arXiv",
    month = "12",
    year = "2001"
}

@article{Compere:2008us,
    author = "Compere, Geoffrey and Marolf, Donald",
    title = "{Setting the boundary free in AdS/CFT}",
    eprint = "0805.1902",
    archivePrefix = "arXiv",
    primaryClass = "hep-th",
    doi = "10.1088/0264-9381/25/19/195014",
    journal = "Class. Quant. Grav.",
    volume = "25",
    pages = "195014",
    year = "2008"
}

@article{Hertog:2005hu,
    author = "Hertog, Thomas and Horowitz, Gary T.",
    title = "{Holographic description of AdS cosmologies}",
    eprint = "hep-th/0503071",
    archivePrefix = "arXiv",
    doi = "10.1088/1126-6708/2005/04/005",
    journal = "JHEP",
    volume = "04",
    pages = "005",
    year = "2005"
}

@article{Duff:1986hr,
    author = "Duff, M. J. and Nilsson, B. E. W. and Pope, C. N.",
    title = "{Kaluza-Klein Supergravity}",
    doi = "10.1016/0370-1573(86)90163-8",
    journal = "Phys. Rept.",
    volume = "130",
    pages = "1--142",
    year = "1986"
}

@book{Green:1987sp,
    author = "Green, Michael B. and Schwarz, J. H. and Witten, Edward",
    title = "{SUPERSTRING THEORY. VOL. 1: INTRODUCTION}",
    isbn = "978-0-521-35752-4",
    series = "Cambridge Monographs on Mathematical Physics",
    publisher = "Cambridge University Press",
    month = "7",
    year = "1988"
}

@article{Candelas:1985en,
    author = "Candelas, P. and Horowitz, Gary T. and Strominger, Andrew and Witten, Edward",
    title = "{Vacuum configurations for superstrings}",
    reportNumber = "NSF-ITP-84-170",
    doi = "10.1016/0550-3213(85)90602-9",
    journal = "Nucl. Phys. B",
    volume = "258",
    pages = "46--74",
    year = "1985"
}

@article{Lu:1998xt,
    author = "Lu, Hong and Pope, C. N. and Stelle, K. S.",
    title = "{M theory / heterotic duality: A Kaluza-Klein perspective}",
    eprint = "hep-th/9810159",
    archivePrefix = "arXiv",
    reportNumber = "CERN-TH-98-303, CTP-TAMU-36-98, IMPERIAL-TP-97-98-77, LPTENS-98-39, UPR-0819-T",
    doi = "10.1016/S0550-3213(99)00086-3",
    journal = "Nucl. Phys. B",
    volume = "548",
    pages = "87--138",
    year = "1999"
}

@article{Freedman:1999gp,
    author = "Freedman, D. Z. and Gubser, S. S. and Pilch, K. and Warner, N. P.",
    title = "{Renormalization group flows from holography supersymmetry and a c theorem}",
    eprint = "hep-th/9904017",
    archivePrefix = "arXiv",
    reportNumber = "CERN-TH-99-86, HUTP-99-A015, USC-99-1, MIT-CTP-2846",
    doi = "10.4310/ATMP.1999.v3.n2.a7",
    journal = "Adv. Theor. Math. Phys.",
    volume = "3",
    pages = "363--417",
    year = "1999"
}

@article{DeWolfe:1999cp,
    author = "DeWolfe, O. and Freedman, D. Z. and Gubser, S. S. and Karch, A.",
    title = "{Modeling the fifth-dimension with scalars and gravity}",
    eprint = "hep-th/9909134",
    archivePrefix = "arXiv",
    reportNumber = "HUTP-99-A048, MIT-CTP-2903",
    doi = "10.1103/PhysRevD.62.046008",
    journal = "Phys. Rev. D",
    volume = "62",
    pages = "046008",
    year = "2000"
}

@article{Ceresole:2007wx,
    author = "Ceresole, Anna and Dall'Agata, Gianguido",
    title = "{Flow Equations for Non-BPS Extremal Black Holes}",
    eprint = "hep-th/0702088",
    archivePrefix = "arXiv",
    doi = "10.1088/1126-6708/2007/03/110",
    journal = "JHEP",
    volume = "03",
    pages = "110",
    year = "2007"
}

@article{Skenderis:2006jq,
    author = "Skenderis, Kostas and Townsend, Paul K.",
    title = "{Hidden supersymmetry of domain walls and cosmologies}",
    eprint = "hep-th/0602260",
    archivePrefix = "arXiv",
    reportNumber = "DAMTP-2006-18, ITFA-2006-08",
    doi = "10.1103/PhysRevLett.96.191301",
    journal = "Phys. Rev. Lett.",
    volume = "96",
    pages = "191301",
    year = "2006"
}

@article{Townsend:1984iu,
    author = "Townsend, P. K.",
    title = "{Positive Energy and the Scalar Potential in Higher Dimensional (Super)gravity Theories}",
    reportNumber = "PRINT-84-0914 (CAMBRIDGE)",
    doi = "10.1016/0370-2693(84)91610-1",
    journal = "Phys. Lett. B",
    volume = "148",
    pages = "55--59",
    year = "1984"
}

@article{Skenderis:1999mm,
  author        = "Skenderis, Kostas and Townsend, Paul K.",
  title         = "{Gravitational Stability and Renormalization-Group Flow}",
  eprint        = "hep-th/9909070",
  archivePrefix = "arXiv",
  journal       = "Phys. Lett. B",
  volume        = "468",
  pages         = "46--51",
  year          = "1999"
}

@article{Breitenlohner:1982bm,
    author = "Breitenlohner, Peter and Freedman, Daniel Z.",
    title = "{Positive Energy in anti-De Sitter Backgrounds and Gauged Extended Supergravity}",
    reportNumber = "PRINT-82-0420 (MIT)",
    doi = "10.1016/0370-2693(82)90643-8",
    journal = "Phys. Lett. B",
    volume = "115",
    pages = "197--201",
    year = "1982"
}

@article{Breitenlohner:1982jf,
    author = "Breitenlohner, Peter and Freedman, Daniel Z.",
    title = "{Stability in Gauged Extended Supergravity}",
    reportNumber = "Print-82-0500 (MIT)",
    doi = "10.1016/0003-4916(82)90116-6",
    journal = "Annals Phys.",
    volume = "144",
    pages = "249",
    year = "1982"
}

@article{Gursoy:2008za,
    author = "Gursoy, U. and Kiritsis, E. and Mazzanti, L. and Nitti, F.",
    title = "{Holography and Thermodynamics of 5D Dilaton-gravity}",
    eprint = "0812.0792",
    archivePrefix = "arXiv",
    primaryClass = "hep-th",
    reportNumber = "CPHT-RR088-1108, SPIN-08-57, ITP-UU-08-74",
    doi = "10.1088/1126-6708/2009/05/033",
    journal = "JHEP",
    volume = "05",
    pages = "033",
    year = "2009"
}

@article{Anabalon:2015xvl,
    author = "Anabalon, Andres and Astefanesei, Dumitru and Choque, David and Martinez, Cristian",
    title = "{Trace Anomaly and Counterterms in Designer Gravity}",
    eprint = "1511.08759",
    archivePrefix = "arXiv",
    primaryClass = "hep-th",
    doi = "10.1007/JHEP03(2016)117",
    journal = "JHEP",
    volume = "03",
    pages = "117",
    year = "2016"
}

@article{Henneaux:2006hk,
    author = "Henneaux, Marc and Martinez, Cristian and Troncoso, Ricardo and Zanelli, Jorge",
    title = "{Asymptotic behavior and Hamiltonian analysis of anti-de Sitter gravity coupled to scalar fields}",
    eprint = "hep-th/0603185",
    archivePrefix = "arXiv",
    reportNumber = "CECS-PHY-06-04, NSF-KITP-06-20, ULB-TH-06-05",
    doi = "10.1016/j.aop.2006.05.002",
    journal = "Annals Phys.",
    volume = "322",
    pages = "824--848",
    year = "2007"
}

@article{Boucher:1984yx,
    author = "Boucher, W.",
    title = "{POSITIVE ENERGY WITHOUT SUPERSYMMETRY}",
    reportNumber = "Print-84-0283 (CAMBRIDGE)",
    doi = "10.1016/0550-3213(84)90394-8",
    journal = "Nucl. Phys. B",
    volume = "242",
    pages = "282--296",
    year = "1984"
}

@article{Faulkner:2010fh,
    author = "Faulkner, Thomas and Horowitz, Gary T. and Roberts, Matthew M.",
    title = "{New stability results for Einstein scalar gravity}",
    eprint = "1006.2387",
    archivePrefix = "arXiv",
    primaryClass = "hep-th",
    doi = "10.1088/0264-9381/27/20/205007",
    journal = "Class. Quant. Grav.",
    volume = "27",
    pages = "205007",
    year = "2010"
}

@article{deHaro:2000vlm,
    author = "de Haro, Sebastian and Solodukhin, Sergey N. and Skenderis, Kostas",
    title = "{Holographic reconstruction of space-time and renormalization in the AdS / CFT correspondence}",
    eprint = "hep-th/0002230",
    archivePrefix = "arXiv",
    reportNumber = "SPIN-2000-05, ITP-UU-00-03, PUTP-1921",
    doi = "10.1007/s002200100381",
    journal = "Commun. Math. Phys.",
    volume = "217",
    pages = "595--622",
    year = "2001"
}

@article{Batrachenko:2004fd,
	author = "Batrachenko, A. and Liu, James T. and McNees, R. and Sabra, W. A. and Wen, W. Y.",
	title = "{Black hole mass and Hamilton-Jacobi counterterms}",
	eprint = "hep-th/0408205",
	archivePrefix = "arXiv",
	reportNumber = "CAMS-04-03, MCTP-04-49",
	doi = "10.1088/1126-6708/2005/05/034",
	journal = "JHEP",
	volume = "05",
	pages = "034",
	year = "2005"
}

@article{Anabalon:2025sqr,
    author = "Anabal{\`o}n, Andr{\`e}s and Astefanesei, Dumitru and Choque, David and Gallerati, Antonio",
    title = "{Thermal superpotential and thermodynamics of neutral hairy black holes in extended SUGRA}",
    eprint = "2511.04428",
    archivePrefix = "arXiv",
    primaryClass = "hep-th",
    month = "11",
    year = "2025"
}

@article{Brown:1992br,
    author = "Brown, J. David and York, Jr., James W.",
    title = "{Quasilocal energy and conserved charges derived from the gravitational action}",
    eprint = "gr-qc/9209012",
    archivePrefix = "arXiv",
    reportNumber = "IFP-423-UNC, TAR-009-UNC",
    doi = "10.1103/PhysRevD.47.1407",
    journal = "Phys. Rev. D",
    volume = "47",
    pages = "1407--1419",
    year = "1993"
}

@article{Balasubramanian:1999re,
    author = "Balasubramanian, Vijay and Kraus, Per",
    title = "{A Stress tensor for Anti-de Sitter gravity}",
    eprint = "hep-th/9902121",
    archivePrefix = "arXiv",
    reportNumber = "HUTP-99-A002, EFI-99-6, NSF-ITP-98-132",
    doi = "10.1007/s002200050764",
    journal = "Commun. Math. Phys.",
    volume = "208",
    pages = "413--428",
    year = "1999"
}

@article{Anabalon:2020qux,
    author = "Anabal\'on, A. and Astefanesei, D. and Choque, D. and Gallerati, A. and Trigiante, M.",
    title = "{Exact holographic RG flows in extended SUGRA}",
    eprint = "2012.01289",
    archivePrefix = "arXiv",
    primaryClass = "hep-th",
    doi = "10.1007/JHEP04(2021)053",
    journal = "JHEP",
    volume = "04",
    pages = "053",
    year = "2021"
}

@article{Myers:1999psa,
    author = "Myers, Robert C.",
    title = "{Stress tensors and Casimir energies in the AdS / CFT correspondence}",
    eprint = "hep-th/9903203",
    archivePrefix = "arXiv",
    reportNumber = "MCGILL-99-06",
    doi = "10.1103/PhysRevD.60.046002",
    journal = "Phys. Rev. D",
    volume = "60",
    pages = "046002",
    year = "1999"
}

@article{Henneaux:1985tv,
    author = "Henneaux, M. and Teitelboim, C.",
    title = "{Asymptotically anti-De Sitter Spaces}",
    doi = "10.1007/BF01205790",
    journal = "Commun. Math. Phys.",
    volume = "98",
    pages = "391--424",
    year = "1985"
}

@article{Anabalon:2014fla,
    author = "Anabalon, Andres and Astefanesei, Dumitru and Martinez, Cristian",
    title = "{Mass of asymptotically anti\textendash{}de Sitter hairy spacetimes}",
    eprint = "1407.3296",
    archivePrefix = "arXiv",
    primaryClass = "hep-th",
    doi = "10.1103/PhysRevD.91.041501",
    journal = "Phys. Rev. D",
    volume = "91",
    number = "4",
    pages = "041501",
    year = "2015"
}

@article{Faedo:2015jqa,
    author = "Faedo, Federico and Klemm, Dietmar and Nozawa, Masato",
    title = "{Hairy black holes in $\rm{N} = 2$ gauged supergravity}",
    eprint = "1505.02986",
    archivePrefix = "arXiv",
    primaryClass = "hep-th",
    reportNumber = "IFUM-1040-FT",
    doi = "10.1007/JHEP11(2015)045",
    journal = "JHEP",
    volume = "11",
    pages = "045",
    year = "2015"
}

@inproceedings{Faedo:2017veq,
    author = "Faedo, Federico and Klemm, Dietmar and Nozawa, Masato",
    title = "{Hairy black holes in $N = 2$ gauged supergravity}",
    booktitle = "{14th Marcel Grossmann Meeting on Recent Developments in Theoretical and Experimental General Relativity, Astrophysics, and Relativistic Field Theories}",
    doi = "10.1142/9789813226609_0562",
    volume = "4",
    pages = "4204--4207",
    year = "2017"
}

@article{Anabalon:2017yhv,
    author = "Anabal\'on, Andr\'es and Astefanesei, Dumitru and Gallerati, Antonio and Trigiante, Mario",
    title = "{Hairy Black Holes and Duality in an Extended Supergravity Model}",
    eprint = "1712.06971",
    archivePrefix = "arXiv",
    primaryClass = "hep-th",
    doi = "10.1007/JHEP04(2018)058",
    journal = "JHEP",
    volume = "04",
    pages = "058",
    year = "2018"
}

@article{Canfora:2021nca,
    author = "Canfora, Fabrizio and Oliva, Julio and Oyarzo, Marcelo",
    title = "{New BPS solitons in $ \mathcal{N} $ = 4 gauged supergravity and black holes in Einstein-Yang-Mills-dilaton theory}",
    eprint = "2111.11915",
    archivePrefix = "arXiv",
    primaryClass = "hep-th",
    doi = "10.1007/JHEP02(2022)057",
    journal = "JHEP",
    volume = "02",
    pages = "057",
    year = "2022"
}

\end{document}